\documentclass[10pt]{article}
\usepackage{amsmath,amssymb,cite,epsfig}
\numberwithin{equation}{section}

\def \beq{\begin{equation}}
\def \eeq{\end{equation}}
\def \beqa{\begin{eqnarray}}
\def \eeqa{\end{eqnarray}}

\def \e{\varepsilon}
\def \l{\left(}
\def \r{\right)}

\begin{document}

\title{\begin{flushright}
\end{flushright}
{\bf Thermodynamics of Ideal Gas in Doubly Special Relativity}}
\author{Nitin Chandra \footnote{nitin@cts.iisc.ernet.in} \,and Sandeep Chatterjee\footnote{sandeep@cts.iisc.ernet.in}\\ 
\begin{small}
Centre for High Energy Physics, Indian Institute of Science, Bangalore, 560012, India.
\end{small}}
\date{\empty}
\maketitle

\begin{abstract}
We study thermodynamics of an ideal gas in Doubly Special Relativity.
New type of special functions (which we call Incomplete Modified Bessel functions) emerge.
We obtain a series solution for the partition function and derive thermodynamic quantities.
We observe that DSR thermodynamics is non-perturbative in the SR and massless limits.
A stiffer equation of state is found.

\end{abstract}

\section{Introduction}

Attempts to combine gravity with quantum mechanics in search of the theory 
for quantum gravity always seem to give rise to the Planck length 
$\left(l_P =\sqrt{\frac{G c}{h^3}}\right)$ 
that provides the scale at which the quantum effects of gravity will show up \cite{Mead:1964zz, Padmanabhan:1986ny, Padmanabhan:1985jq, Veneziano:1986zf, Amati:1988tn, Konishi:1989wk, Greensite:1990jm, Maggiore:1993rv, Garay:1994en, Rovelli:1994ge, AmelinoCamelia:2002vw}.
The existence of such a length scale is in conflict with the equivalence principle
because observers in different inertial frames will not agree on $l_P$ due to the 
Lorentz-Fitzgerald contraction. It has been shown that it is possible to still have equivalence 
principle by deforming Special Relativity (SR). These classes of theories fall under 
the name Doubly Special Relativity (DSR) \cite{AmelinoCamelia:2000ge, AmelinoCamelia:2000mn, AmelinoCamelia:2002vy, AmelinoCamelia:2002wr}. In DSR, apart from the constancy of speed-of-light scale, the Planck length $l_P$ or equivalently Planck energy $\kappa$ is also 
constant under coordinate transformation from one inertial frame to another. This leads to modification in  the dispersion relation. Consequences of the modified dispersion relations on the thermodynamics are being studied extensively to infer the effect of Planck scale physics \cite{AmelinoCamelia:2004xx, Camacho:2006qg, Gregg:2008jb, AmelinoCamelia:2009tv, Camacho:2007qy, Alexander:2001ck, Bertolami:2009wa}. The effect of modified dispersion relations in loop-quantum-gravity on black hole thermodynamics was studied in \cite{AmelinoCamelia:2004xx}. The same as a Lorentz violating phenomena on the thermodynamics of macroscopic systems (like white dwarfs) \cite {Camacho:2006qg, Gregg:2008jb, AmelinoCamelia:2009tv} and as a noncommutative phenomena on cosmology and astrophysical systems \cite{Alexander:2001ck, Bertolami:2009wa} have also been studied. Moreover, photon gas thermodynamics in the context of modified dispersion relations \cite{Camacho:2007qy} and DSR \cite{Das:2010gk} are being investigated. In \cite{Das:2010gk} the effect comes solely because of the presence of a maximum energy scale as the photon dispersion relation remains unmodified.

The present paper aims to study the thermodynamics of an ideal gas consisting of massive particles in DSR scenario. Both the modification in the dispersion relation of the constituent particles and the presence of a maximum energy scale are expected to contribute to new effects. DSR transformations can be of several type. In this paper, we follow
the formulation of \cite{Magueijo:2001cr, Magueijo:2002am} where the modified dispersion relation becomes
\begin{equation}
\e^2-p^2=m^2\l1-\frac{\e}{\kappa}\r^2.
\label{MS}
\end{equation}
As  $0\leq \l 1-\frac{\e}{\kappa}\r^{2} \leq 1$, the energy of a particle with a given momentum decreases in DSR. This has consequence on the thermodynamics as we will see in $\mathsection$ \ref{thermodynamics}.
The parameter $m$ can be called ``invariant mass'' as it remains invariant under a DSR transformation. 
Note that in contrary to the SR case, $m$ is no more the rest mass energy of the particle.
To get the rest mass energy $m_0$, we put $p=0$ in (\ref{MS}). We get two expressions 
for $m_0$, namely
\begin{equation}
m_0=\frac{\pm m}{1\pm \frac{m}{\kappa}}.
\end{equation}
The two solutions are connected by the redefinition of the parameter $m\rightarrow -m$.
Henceforth, without any loss of generality we use
\begin{equation}
m_0=\frac{m}{1+\frac{m}{\kappa}}.
\end{equation}
The physical world is characterized by $E<\kappa$ \cite{Magueijo:2002am}.
In this sub-Planck regime ($E_{p=0}=m_0<\kappa$), the positivity of rest mass ($m_0\geq 0$) 
restricts the range of the invariant mass to $0\leq m<\infty$.
Thus, in (\ref{MS}), we have $0\leq p,E < \kappa$ and $0\leq m<\infty$.

We study the thermodynamics of an ideal gas in DSR 
setup. We obtain a series solution for the  partition function 
and compute the various thermodynamic quantities.
We show that our results go to the standard results in the 
SR limit ($\kappa \rightarrow \infty$) \cite{greinerbook} as well as in the 
massless DSR limit \cite{Das:2010gk}.
\section{The Partition Function}
We consider a gaseous system of non-interacting particles obeying Maxwell-Boltzmann
statistics whose macrostate is denoted by $(N,V,T)$ where $N$ is the number of particles
 in the system confined in volume $V$ at a temperature $T$. In the canonical
ensemble the thermodynamics of this system is derived from its partition function \cite{pathria}
\begin{equation}
 Z_N\l V,T\r=\sum_E\exp[-\beta E],
\end{equation}
where $\beta={1\over k_BT}$ and $\displaystyle{\sum_{E}}$ denotes sum over all the energy eigenvalues 
of the system. The total energy $E$ of the system can be written in terms of single particle energy $\e$
\beq
E=\sum_{\e}n_{\e}\e,
\eeq
where $n_{\e}$ is the number of particles in the single-particle energy state $\e$ and satisfy the following condition
\beqa
\sum_{\e}n_{\e}&=&N.
\label{cond}
\eeqa
We can rewrite $Z_N$ as
\begin{equation}
 Z_N\l V,T\r={\sum_{\{n_{\e}\}}}^\prime g\{n_{\e}\}\exp[-\beta\sum_{\e}n_{\e}\e],
\end{equation}
where $g\{n_{\e}\}$ is the statistical weight factor appropriate to the distribution set $\{n_{\e}\}$. 
The summation ${\sum}^{\prime}$ goes over all distribution sets
that conform to the above restrictive condition (\ref{cond}).
For Maxwell-Boltzmann statistics, it can be shown
\beq
Z_N\l V,T\r={1\over N!}[Z_1\l V,T\r]^N,
\eeq
where $Z_1$ is the single particle partition function given by
\beq
Z_1\l V,T\r=\sum_{\e}\exp[-\beta\e].
\label{Zsingle}
\eeq
While for ordinary spacetime, it is easy to show that in the large volume limit one can replace the sum by an integral
\beq
\sum_{\e} \rightarrow {V\over h^3}\int d^3p,
\label{replace}
\eeq
for more exotic spacetimes the measure of integration is expected to get modified
\beq
d^3p \rightarrow f(\vec{p}) d^3p.
\label{modmeas}
\eeq
Hence putting together (\ref{Zsingle}), (\ref{replace}), (\ref{modmeas}) and taking $\hbar = k_B = 1$ we get
\beqa
Z_1\l V,T\r &=& \frac{V}{\l2\pi\r^3}\int_{p=0}^{\kappa}d^3p\,\, f(\vec{p}) \exp[-\beta \l\e-m_0\r].
\label{Z1_modmeas}
\eeqa
Note that in accordance with standard practice, we have subtracted the rest mass $m_0$ from the
relativistic energy $\e$ of the particle.
Although there have been few attempts\cite{AmelinoCamelia:2009tv, AmelinoCamelia:1999pm, KowalskiGlikman:2001ct}, the form of $f(\vec{p})$ is far from settled.
Assuming isotropy of spacetime we may take $f(\vec{p})=f(p)$. For a possible deformation of the integration measure, $f(p)$ should be expandable in Taylor series in $\frac{\e}{\kappa}$
\begin{equation}
f(p) = \displaystyle{\sum_{n=0}^{\infty}} \frac{a_{n}}{n!}\left(\frac{\e}{\kappa}\right)^{n},
\end{equation}
with $a_{0}=1$ since in the limit $\kappa\rightarrow\infty$, $f(p)\rightarrow 1$. Hence $Z_1(V,T)$ becomes
\beqa
Z_1\l V,T\r &=& \frac{V}{\l2\pi\r^3}\int_{p=0}^{\kappa}d^3p\,\, \sum_{n=0}^{\infty} \frac{a_{n}}{n!}\left(\frac{\e}{\kappa}\right)^{n} \exp[-\beta \l\e-m_0\r]
\label{Z1_final1}\\
&=&\displaystyle{\sum_{n=0}^{\infty}} \frac{a_{n}}{n! \kappa^{n}} \l m_{0}-\frac{\partial}{\partial \beta}\r^{n} Z_1^0\l V,T\r,
\label{Z1_final2}
\eeqa
where $Z_1^0\l V,T\r$ is the single particle partition function with unmodified measure
\beq
Z_1^0\l V,T\r=\frac{V}{\l2\pi\r^3}\int_{p=0}^{\kappa}d^3p \exp[-\beta \l\e-m_0\r].
\label{z10}
\eeq
The derivation of (\ref{Z1_final2}) from (\ref{Z1_final1}) involves two steps. Firstly, the interchange of the summation
and integration which is allowed if (see theorem 1.38 of \cite{rudin})
\beqa
\displaystyle{\sum_{n=0}^{\infty}} \frac{|a_{n}|}{n! \kappa^{n}} \int_{p=0}^{\kappa}d^3p\,\,  \e^{n} \exp[-\beta \l\e-m_0\r]
&=&\displaystyle{\sum_{n=0}^{\infty}} \frac{|a_{n}|}{n! \kappa^{n}} \l m_{0}-\frac{\partial}{\partial \beta}\r^{n} Z_1^0\l V,T\r < \infty.
\label{condition_modmeas_expansion}
\eeqa
Secondly, writing $\int_{p=0}^{\kappa}d^3p\,\,  \e^{n} \exp[-\beta \l\e-m_0\r]$ as $\l m_{0}-\frac{\partial}{\partial \beta}\r^{n} \int_{p=0}^{\kappa}d^3p\,\,  \exp[-\beta \l\e-m_0\r]$ since the former can be written as $4\pi \int_{p=0}^{\kappa}dp\,\, p^{2} \e^{n} \exp[-\beta \l\e-m_0\r] $ whose integrand $p^{2} \e^{n} \exp[-\beta \l\e-m_0\r]$ remains to be continuous and bounded for $p\in[0,\kappa], \beta\in [0,\infty]$ (see $\S$ 5.12 of \cite{fleming}).
Hence our problem has boiled down to solving the integral in (\ref{z10})
where $\e$ and $p$ are related by the modified DSR dispersion relation given in (\ref{MS}). The solution of 
$Z_1^0$ in the massless case obtained in \cite{Das:2010gk} is
\beq
Z_{1ml}^0=\frac{2V}{(2\pi)^2\beta^3}\left(2-e^{-\beta\kappa}(\beta^2\kappa^2+2\beta\kappa+2)\right).
\label{DSRphoton}
\eeq
The term with $\displaystyle{e^{-\beta\kappa}}$ makes $Z_{1ml}^0$ non-analytic at $\frac{1}{\kappa}=0$. We anticipate
that even when $m_0\neq 0$, $Z_1^0$ continues to be non-analytic at $\frac{1}{\kappa}=0$ and hence does not
admit a straightforward Taylor series expansion in $\frac{1}{\kappa}$. Thus in order to find the leading order deviation of DSR thermodynamics 
from the SR case, one would require a non-trivial series expansion.
\subsection{Solution of $Z_1^0$}
Changing the variable from $p$ to $\e$ in (\ref{z10}) we get
\begin{equation}
Z_1^0\l V, \beta\r = \frac{2V}{\l2\pi\r^2}\exp[\beta m_0]\int_{m_0}^{\kappa}\left[\e+
\frac{m^2}{\kappa}\l1-\frac{\e}{\kappa}\r\right]
\left[\e^2-m^2\l1-\frac{\e}{\kappa}\r^2\right]^{1/2}\exp[-\beta \e]d\e.
\label{Z_E}
\end{equation}
We now consider three different regions of values of $m$:

\subsubsection{Case I: $m=\kappa$}
In this case the partition function reduces to
\begin{eqnarray}
Z_1^0\l V,T\r &=& \frac{2V}{\l2\pi\r^2}\kappa^{3/2}\exp\left[{\beta\kappa\over2}\right]\int_{\kappa/2}^{\kappa}d\e(2\e-\kappa)^{1/2}\exp\left[-\beta \e\right]\nonumber\\
& = &\frac{2\sqrt 2V}{\l2\pi\r^2}\l\frac{\kappa}{\beta}\r^{3/2}\gamma\l\frac{3}{2},\frac{\beta\kappa}{2}\r,
\label{Zcase2}
\end{eqnarray}
where $\gamma(a,x)$ is the Incomplete Gamma Function (see (6.5.2) of 
\cite{abramowitz}). The factor $\left[\e+\frac{m^2}{\kappa}\l1-\frac{\e}{\kappa}\r\right]$ reduces to $\kappa$ and this simplifies the integral in (\ref{Z_E}) yielding a simple analytical form for $Z^0_1(V,T)$. 

\subsubsection{Case II: $\kappa<m<\infty$}
We put $t=\frac{\e}{m}\left[\left(\frac{m}{\kappa}\right)^2-1\right]-\frac{m}{\kappa}$ 
in (\ref{Z_E}) to get 
\begin{eqnarray}
Z_1^0 &=& -\frac{2Vm^3}{\l2\pi\r^2\left[\l\frac{m}{\kappa}\r^2-1\right]^{3/2}}\exp\left[\beta m_0-\frac{\beta^{\prime}m^2}{\kappa}\right]
\int_{-1}^{-\kappa /m}dt \,\,\,\,\,\,t (1-t^2)^{1/2}\exp\left[-\beta^{\prime}mt\right]\nonumber\\
&=& -\frac{2Vm^3}{\l2\pi\r^2\left[\l\frac{m}{\kappa}\r^2-1\right]^{3/2}}\exp\left[\beta m_0-\frac{\beta^{\prime}m^2}{\kappa}\right]
\left[I^*\left(\beta^{\prime}m,1\right)-I^*\left(\beta^{\prime}m,\frac{\kappa}{m}\right)\right],
\label{Istar}
\end{eqnarray}
where $\beta^{\prime}=\frac{\beta}{\l\frac{m}{\kappa}\r^2-1}$ and
\begin{equation}
I^*(x,y) = \int_{-y}^{1}dt \,\,\,\,\,\,t (1-t^2)^{1/2}\exp\left[-xt\right].
\label{istar}
\end{equation}
We define Incomplete Modified Bessel function $I_{\nu}(z,y)$ of order $\nu$
\beq
I_{\nu}(z,y)=\frac{1}{\sqrt{\pi}\Gamma(\nu+\frac{1}{2})}\l\frac{z}{2}\r^{\nu}
\int_{-y}^1(1-t^2)^{\nu-\frac{1}{2}}\exp[-zt]dt\quad [Re\,\,\nu>0,|arg \,\,z|<\frac{\pi}{2}],
\eeq
such that $I^*(x,y)=-\frac{\partial}{\partial x}[\frac{\pi}{x}I_1(x,y)]$. In particular for $y=1$, 
using (3.387 (1)) of \cite{gradshteyn} and (9.6.26) of \cite{abramowitz} we get
\begin{equation}
I^*(x,1) = -\frac{\pi I_2(x)}{x},
\label{istar_i2}
\end{equation}
where $I_2(x)$ is the 2nd order Modified Bessel function. In the limit $m\rightarrow\infty$ $(m_0\rightarrow \kappa)$ 
one gets from (\ref{MS})\footnote[1]{We would like to thank Diptiman Sen for pointing out this interesting case.}
\beq
\frac{\e^2-p^2}{m^2}=\l1-\frac{\e}{\kappa}\r^2\Rightarrow \e\rightarrow\kappa \quad \forall p\in[0,\kappa].\nonumber
\eeq
Thus the total energy $E$ of the system becomes $E=N\kappa$ and the thermodynamics simplifies.
Entropy can be computed by counting the total number of microstates $\Omega_N$ available to the system
\beq
\Omega_N=\frac{\Omega_1^N}{N!}=\frac{1}{N!}\l V\int_{p=0}^{\kappa}\frac{d^3p}{h^3}\r^N=\frac{1}{N!}\l\frac{2V\kappa^3}{3\l2\pi\r^2}\r^N,
\eeq
where $\Omega_1$ is the total number of microstates available for a single particle. Thus the entropy $S$ of the system is
\beq
S=\ln\left[\frac{1}{N!}\l\frac{2V\kappa^3}{3\l2\pi\r^2}\r^N\right].
\label{s}
\eeq
The first law of thermodynamics in this case becomes 
\beq
dE=-PdV+\mu dN.
\eeq
Note that the usual term $TdS$ has been dropped as from (\ref{s}) it is evident that $S$ is a function of $N$ and $V$ alone.
The pressure of the system is zero as $P=-\left.\frac{\partial E}{\partial V}\right|_{N}=0$ while the chemical potential
is $\mu=\left.\frac{\partial E}{\partial N}\right|_{V}=\kappa$.

Equation (\ref{z10}) can now be easily integrated to give
\beq
Z_1^0=\frac{2V}{\l2\pi\r^2}\frac{\kappa^3}{3},
\label{istarminf}
\eeq
which gives the limiting behaviour of $I^*\l\beta^{\prime}m,\frac{\kappa}{m}\r$ using (\ref{istar}) and (9.6.7) of \cite{abramowitz}
\beq
I^*\l\frac{\beta m}{\l\frac{m}{\kappa}\r^2-1},\frac{\kappa}{m}\r\stackrel{m\rightarrow\infty}{\longrightarrow}\frac{1}{3}.
\eeq
\subsubsection{Case III: $0<m<\kappa$}
We put $t=\frac{\e}{m}\left[1-\l\frac{m}{\kappa}\r^2\right]+\frac{m}{\kappa}$ 
in (\ref{Z_E}) to get 
\begin{eqnarray}
Z_1^0 &=& \frac{2Vm^3}{\l2\pi\r^2\left[1-\l\frac{m}{\kappa}\r^2\right]^{3/2}}\exp\left[\beta m_0+\frac{\beta^{\prime \prime}m^2}{\kappa}\right]
\int_1^{\kappa /m}dt \,\,\,\,\,\,t (t^2-1)^{1/2}\exp\left[-\beta^{\prime \prime}mt\right]\nonumber\\
&=&\frac{2Vm^3}{\l2\pi\r^2\left[1-\l\frac{m}{\kappa}\r^2\right]^{3/2}}\exp\left[\beta m_0+\frac{\beta^{\prime \prime}m^2}{\kappa}\right]
\left[K^*\left(\beta^{\prime \prime}m,1\right)-K^*\left(\beta^{\prime \prime}m,\frac{\kappa}{m}\right)\right],
\label{Z}
\end{eqnarray}
where $\beta^{\prime \prime}=\frac{\beta}{1-\left(\frac{m}{\kappa}\right)^2}$ and
\begin{equation}
K^*(x,y) = \int_y^{\infty}dt \,\,\,\,\,\,t (t^2-1)^{1/2}\exp\left[-xt\right].
\label{kstar}
\end{equation}
As in Case II, we define Incomplete Modified Bessel function $K_{\nu}(z,y)$ of order $\nu$
\beq
K_{\nu}(z,y)=\frac{\sqrt{\pi}}{\Gamma(\nu+\frac{1}{2})}\l\frac{z}{2}\r^{\nu}\int_{y}^\infty(t^2-1)^{\nu-\frac{1}{2}}\exp[-zt]dt
\quad [Re\,\,\nu>-\frac{1}{2},|arg \,\,z|<\frac{\pi}{2}],
\eeq
such that $K^*(x,y)=-\frac{\partial}{\partial x}[\frac{K_1(x,y)}{x}]$. In particular for $y=1$, using (9.6.23) and (9.6.26) of \cite{abramowitz} we get
\begin{equation}
K^*(x,1) = \frac{K_2(x)}{x},
\label{kstar_k2}
\end{equation}
where $K_2(x)$ is the 2nd order Modified Bessel function.

We shall now obtain the series solution of $K^*(x,y)$.
We rewrite (\ref{kstar}) as
\begin{equation}
K^*\l x,y\r = \int_{y}^{\infty}dt \,\,\,\,\,\,t^2 \l1-\frac{1}{t^2}\r^{1/2}e^{-xt}.
\end{equation}
Inside the integral $t\geq y$ and for $y>1$ (which is a valid assumption for the case of our interest) the factor $\left(1-\frac{1}{t^2}\right)^{1/2}$ can be expanded in series of $\frac{1}{t^2}$ to get
\begin{equation}
K^*\l x,y\r = \int_{y}^{\infty}d\mu_t \,\,\,\,\,\, \left[1+\sum_{r=1}^{\infty}f_r(t)\right]
\label{kstar_series}
\end{equation}
with
\begin{equation}
d\mu_t=t^2e^{-xt}dt 
\end{equation}
and
\begin{equation}
 f_r(t)=\frac{t_r}{t^{2r}},
\end{equation}
where
\begin{equation}
 t_r=\frac{(0-\frac{1}{2})(1-\frac{1}{2})...(r-1-\frac{1}{2})}{r!}=-\frac{(2r-2)!}{2^{2r-1}r!(r-1)!}.
\end{equation}
Now the integral and the summation in (\ref{kstar_series}) can be interchanged if 
$\displaystyle{\sum_{r=1}^{\infty}} \displaystyle{\int_y^{\infty}}d\mu_t|f_r(t)|$ is finite (see theorem 1.38 of \cite{rudin}).
Now as $t_r$ is $-ve$ for all $r\geq 1$ we have $|f_r(t)|=-f_r(t)$.
This allows us to interchange the summation and the integral if the final series is converging. So we get
\begin{equation}
 K^*\l x,y\r = M_0-\frac{1}{2}M_1+\sum_{r=2}^{\infty}t_rM_r,
\label{kstar_mr}
\end{equation}
if the above is a converging series (see Appendix \ref{kstar_convergence} for convergence of $K^*\l x,y\r$). Here
\begin{equation}
 M_r = \int_{y}^{\infty}dt \,\,\,\,\, t^{2(1-r)}e^{-xt}
\end{equation}
for $r=2, 3,...$. \\
$M_0$ and $M_1$ can be easily calculated to be
\begin{equation} \label{m0}
 M_0 = \frac{\exp\l-xy\r}{x^3}\l(xy)^2 + 2xy + 2\r,
\end{equation}
\begin{equation} \label{m1}
 M_1 = \frac{\exp\l-xy\r}{x}.
\end{equation}
Now, changing the variable to $t^\prime=xt$ in $M_r$ for $r\geq 2$ 
we get
\begin{equation}
 M_r = x^{2r-3}\int_{xy}^{\infty}dt^\prime \frac{e^{-t^\prime}}{(t^{\prime})^{2r-2}}.
\end{equation}
Taking $e^{-t^\prime}$ as first function, if we do the integration by parts again and again we finally get
\begin{equation}
 M_r=-\frac{x^{2r-3}}{(2r-3)!}E_1(xy)+e^{-xy}\sum_{k=1}^{2r-3}\frac{(-x)^{k-1}}{(2r-3)(2r-4)...(2r-2-k)}\left(\frac{1}{y}\right)^{2r-2-k}.
\label{mr_series}
\end{equation}
Here $E_1(x)$ is the Exponential Integral (see (5.1.1) of \cite{abramowitz}). A similar attempt to obtain the series 
solution of $I^*(x,y)$ fails. 

Although we obtain the solutions of $Z_1^0$ in three different regions of 
values of $m$, $Z_1^0$ can be shown to be smooth in $m$ (see Appendix \ref{app1}) and hence we do not expect any phase transition like thermodynamic discontinuity as we vary $m$. We use the continuity of $Z_1^0$ to obtain the limiting behaviour of $I^*(\beta^{\prime}m,\frac{\kappa}{m})$ and $K^*(\beta^{\prime}m,\frac{\kappa}{m})$ as $m\rightarrow\kappa$. From (\ref{Zcase2}), (\ref{Istar}), (\ref{istar_i2}) and (9.7.1) of \cite{abramowitz} we obtain the leading order behaviour of $I^*\l\beta^{\prime}m,\frac{\kappa}{m}\r$ as $m \rightarrow \kappa^+$ to be
\beq
I^*\l\beta^{\prime}m,\frac{\kappa}{m}\r \stackrel{m\rightarrow \kappa^+}{\longrightarrow}{2\over (\beta \kappa)^{3/2}}\delta^{3/2}e^{{\beta \kappa\over 2\delta}}\left[-\pi^{1/2}+2e^{-{\beta \kappa\over 2}}\gamma\left({3\over 2}, {\beta \kappa\over 2}\right)\right],
\eeq
with $\delta=\frac{m}{\kappa}-1$.
For $m \rightarrow \kappa^-$, using (\ref{Zcase2}), (\ref{Z}), (\ref{kstar_k2}) and (9.7.2) of \cite{abramowitz} the leading order behaviour of $K^*\l\beta^{\prime\prime}m,{\kappa\over m}\r$ turns out to be
\begin{equation}
K^*(\beta^{\prime \prime}m,{\kappa\over m}) \stackrel{m \rightarrow \kappa^-}{\longrightarrow}{2\over (\beta \kappa)^{3/2}}\epsilon^{3/2}e^{-{\beta \kappa\over 2\epsilon}}\left[\pi^{1/2}-2e^{-{\beta \kappa\over 2}}\gamma\left({3\over 2}, {\beta \kappa\over 2}\right)\right],
\end{equation}
where $\epsilon = 1-{m\over\kappa}$.
\subsection{Leading order deviations}
Having obtained the series solution of $Z_1^0$ in Case III, we shall now obtain the leading order
corrections from the massless and the SR cases.

\subsubsection{Leading order deviation from the massless case}
\begin{figure}
 \begin{center}
  \scalebox{0.75}{\includegraphics{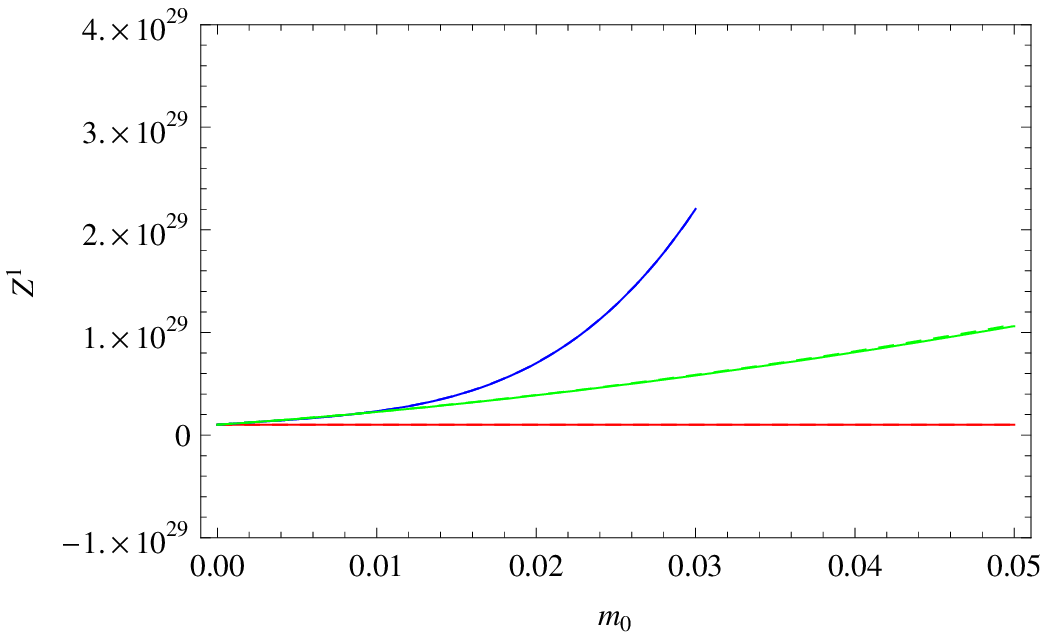}}
  \scalebox{0.75}{\includegraphics{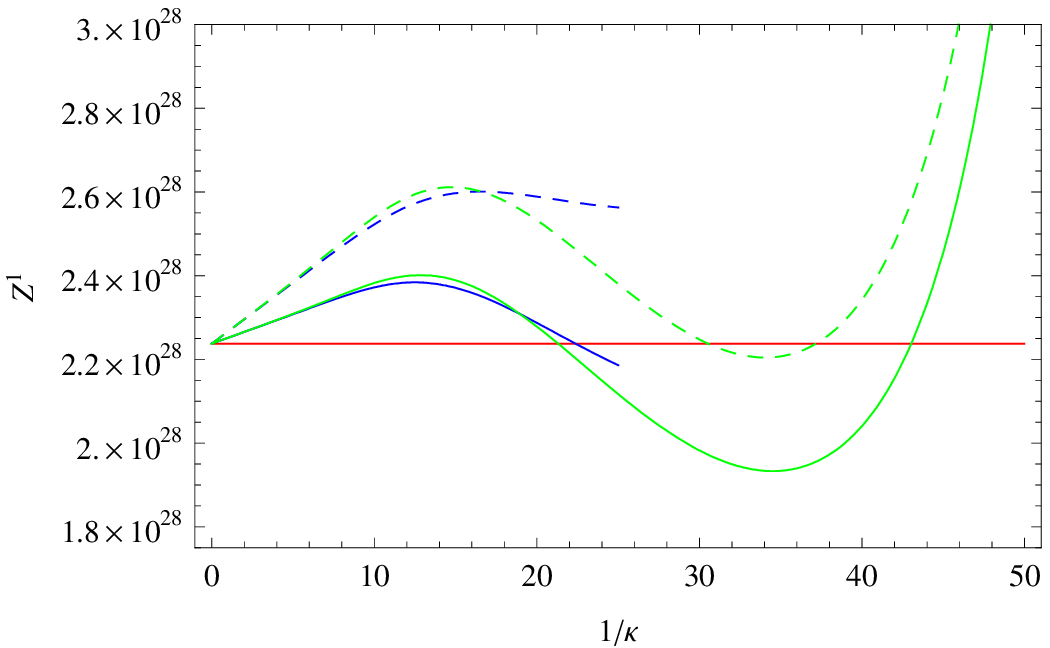}}
\end{center}
\caption{The single particle partition function is plotted vs $m_0$ (left) and ${1\over\kappa}$ (right). The plots for the unmodified measure are in solid lines while those for modified measure are in dashed lines.
The plots in massless (left) and SR (right) cases are shown in red.
The leading order behaviours are plotted in blue. The numerical plots for (\ref{Z}) and (\ref{Z1_final2}) are shown in green (solid and dashed respectively) for comparison.
Different values for the parameters in natural units are as follows: $V=10^{35}, N=10^{25}, T=0.01, a_{0}=1, a_{1}=0.2, a_{2}=a_{3}=...=0$ and $\kappa=1$ (left), $m_0=0.01$ (right).
}
\label{fg.Z}
\end{figure}

Thermodynamics of a photon gas in DSR with dispersion relation (\ref{MS}) and unmodified measure has been worked out in \cite{Das:2010gk}. Here we calculate the deviation of single particle partition function from that of a photon gas. On expanding $Z_1^0$ in $\eta = \frac{m_0}{\kappa}$ with $m_0 \rightarrow 0$ (assuming $\kappa$ to be finite) and using (9.6.10) and (9.6.11) of \cite{abramowitz}, we get
\beq
Z_1^0=Z_{1 ml}^0+Z_{1 ml corr}^0,
\eeq
where $Z_{1 ml}^0$ is the single particle partition function of photon gas in DSR scenario with unmodified measure\cite{Das:2010gk} and $Z_{1 ml corr}^0$ is $\mathcal{O}\l\eta\r$:
\beqa
Z_{1ml}^0&=&\frac{2V}{(2\pi)^2\beta^3}\left(2-e^{-\beta\kappa}(\beta^2\kappa^2+2\beta\kappa+2)\right),\nonumber\\
Z_{1mlcorr}^0&=& - \frac{2V}{(2\pi)^2\beta^3}\frac{(\beta\kappa)^4}{8}\ln(\eta)\l\eta^4+\mathcal{O}(\eta^5)\r+\l\beta\kappa Z_{1ml}^0\r \eta + \mathcal{O}(\eta^2).
\label{zm0}
\eeqa
Note that the correction due to mass of the constituent particle is non-perturbative in nature as the first term in $Z_{1mlcorr}^0$ which contains $\ln(\eta)$ is the non-analytic piece and does not allow a trivial Taylor series expansion at $\eta=0$.
We can rewrite (\ref{Z1_final2}) as
\beq
Z_1=Z_{1 ml}+Z_{1 ml corr},
\eeq
where
\beqa
Z_{1 ml}&=&\displaystyle{\sum_{n=0}^{\infty}} (-1)^{n}\frac{a_{n}}{n!\kappa^{n}}\frac{\partial^{n}Z_{1 ml}^0}{\partial \beta^{n}},\nonumber\\
Z_{1 ml corr}&=&Z^0_{1 ml corr} + \displaystyle{\sum_{n=1}^{\infty}}(-1)^{n}\frac{a_{n}}{n!\kappa^{n}}\l \frac{\partial^{n}Z^0_{1 ml corr}}{\partial\beta^{n}}-\eta n \kappa \frac{\partial^{n-1}Z_{1 ml}^0}{\partial\beta^{n-1}} \r.
\eeqa
The above leading order behaviours have been plotted in Fig \ref{fg.Z}.
For our choice of parameters they match with the numerical plots up to $\frac{m_0}{\kappa}\sim 0.012$. 

\subsubsection{Leading order deviation from the SR case}
On expanding $Z_1^0$ in $\eta = \frac{m_0}{\kappa}$ with $\kappa\rightarrow\infty$ (assuming $m_{0}$ to be finite), we get
\beq
Z_1^0=Z_{1 SR}^0+Z_{1 SR corr}^0,
\eeq
where $Z_{1 SR}^0$ is the single particle partition function in SR and $Z_{1 SR corr}^0$ is $\mathcal{O}\l\eta\r$.
\beqa
Z_{1 SR}^0&=&\frac{2Vm_0^3}{(2\pi)^2}e^{\beta m_0} \frac{K_2(\beta m_0)}{\beta m_0}\nonumber\\
Z_{1 SR corr}^0 &=& - \frac{2Vm_0^3}{(2\pi)^2}\frac{e^{\beta m_0}}{\beta m_0}\frac{e^{-\beta m_0/\eta}}{\eta^2} \l1+\mathcal{O}(\eta)\r 
+\left(1 -\frac{K_1(\beta m_0)}{K_2(\beta m_0)} \right)\beta m_0Z^0_{1SR}\eta
+ \mathcal{O}\left({\eta^2}\right).
\label{sr}
\eeqa
Note that the DSR correction is non-perturbative in nature as the first term in $Z_{1 SR corr}^0$ which contains $e^{-\beta m_0/\eta}$ is the non-analytic piece and does not allow a Taylor series expansion at $\eta=0$. 
This is a novel feature in DSR as we know that SR thermodynamics is perturbative
in the non-relativistic limit:
\beqa
Z_{1SR}^0&=&\frac{4\pi V\l k_BTm_0\r^{3/2}}{h^3}u^{1/2}e^{u}K_2(u)\nonumber\\
&&\stackrel{u\rightarrow \infty}{\longrightarrow}V\l\frac{2\pi m_0k_BT}{h^2}\r^{3/2}\l 1+\frac{15}{8u}+\mathcal{O}\l\l\frac{1}{u}\r^2\r\r\nonumber\\
&&=Z_{1NR}\l 1+\frac{15}{8u}+\mathcal{O}\l\l\frac{1}{u}\r^2\r\r,
\eeqa
where $u=\frac{m_0c^2}{k_BT}$ and $Z_{1NR}$ is the single particle partition function in the non-relativistic case.
We can rewrite (\ref{Z1_final2}) as
\beq
Z_1=Z_{1 SR}+Z_{1 SR corr},
\eeq
where
\beqa
Z_{1 SR}&=&Z_{1 SR}^0,\nonumber\\
Z_{1 SR corr}&=&Z^0_{1 SR corr} + \frac{a_{1}m_0}{\kappa}Z_{1 SR}^0-\frac{a_{1}}{\kappa}\frac{\partial Z^0_{1 SR}}{\partial \beta}.
\eeqa
The above leading order behaviours have been plotted in Fig \ref{fg.Z}.
For our choice of parameters they match with the numerical plots up to $\frac{m_0}{\kappa}\sim 0.08$.
Having obtained the leading order correction to $Z_1$ due to DSR, we shall now compute its
effect on the various thermodynamic quantities. 

\section{Thermodynamic quantities} \label{thermodynamics}
The free energy $F$, pressure $P$, entropy $S$, internal energy $U$, internal energy density $\rho$ and heat capacity $C_{V}$ are defined as
\beqa
F &=& -{1\over \beta}\ln\left(Z_N(V,\beta,m_0)\right) = -\frac{1}{\beta} N\left\{\ln\left({Z_1\over N}\right)+1\right\} \\
P &=& -\left({\partial F\over \partial V}\right)_{N,T} \\
S &=& -\left({\partial F\over \partial T}\right)_{V,N} \\
U &=& F+TS \\
\rho &=& \frac{U}{V} \\
C_V &=& \left({\partial U\over \partial T}\right)_{N,V}
\eeqa 
\begin{figure}
\vspace{-1.75 cm}
 \begin{center}
  \scalebox{0.75}{\includegraphics{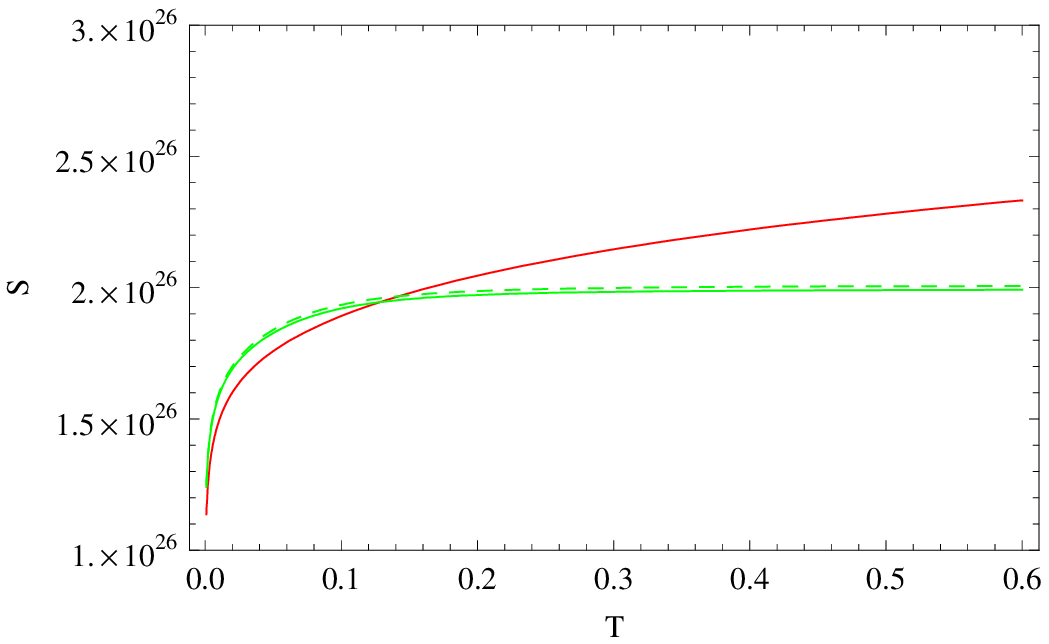}}
  \scalebox{0.75}{\includegraphics{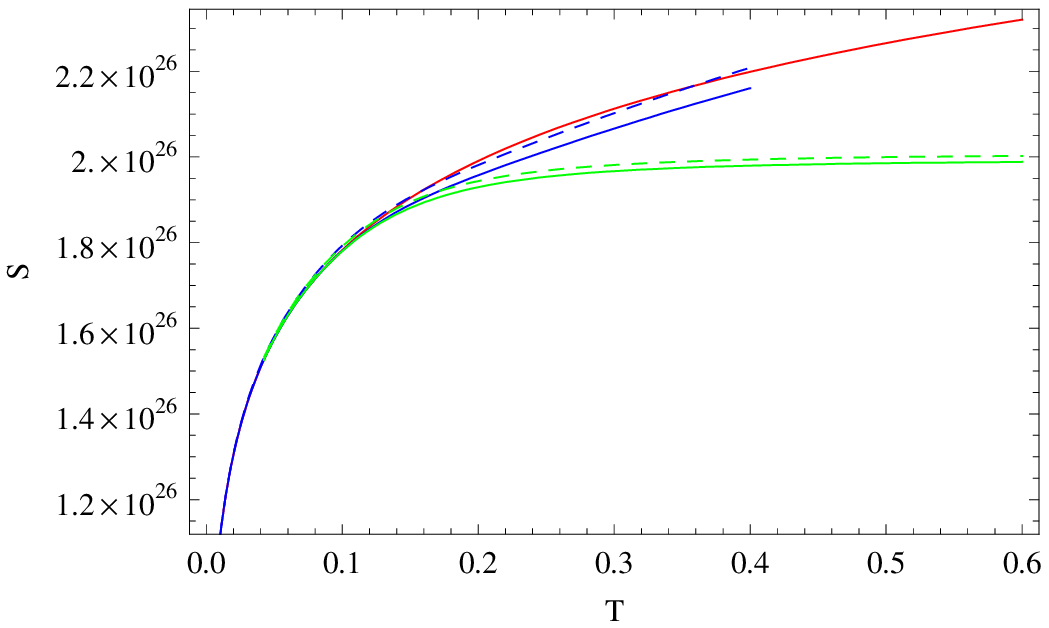}} \\
  \scalebox{0.75}{\includegraphics{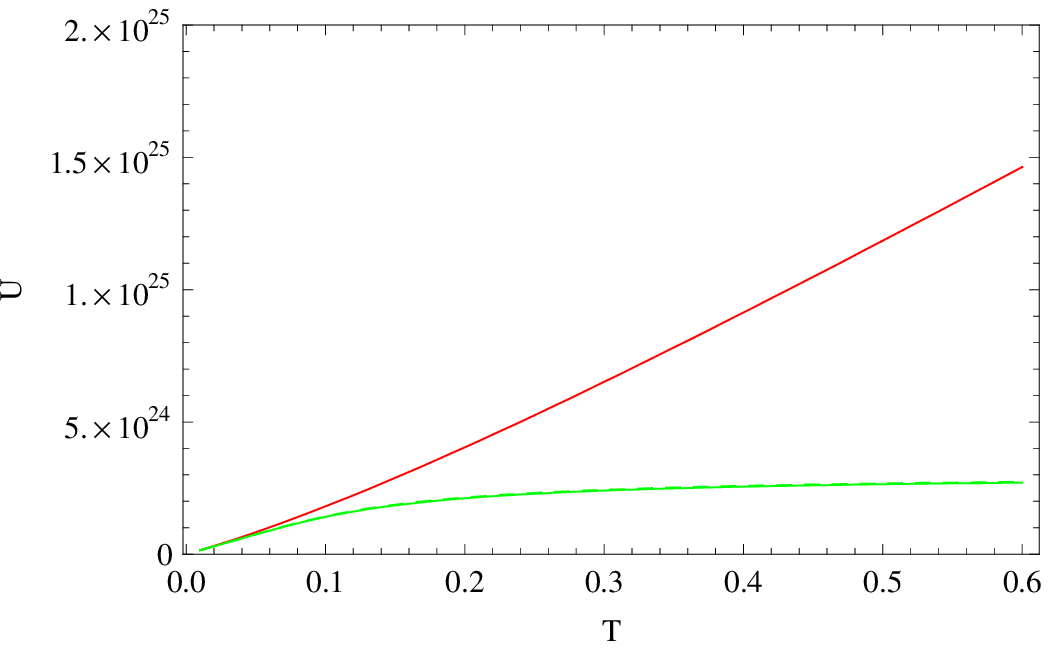}}
  \scalebox{0.75}{\includegraphics{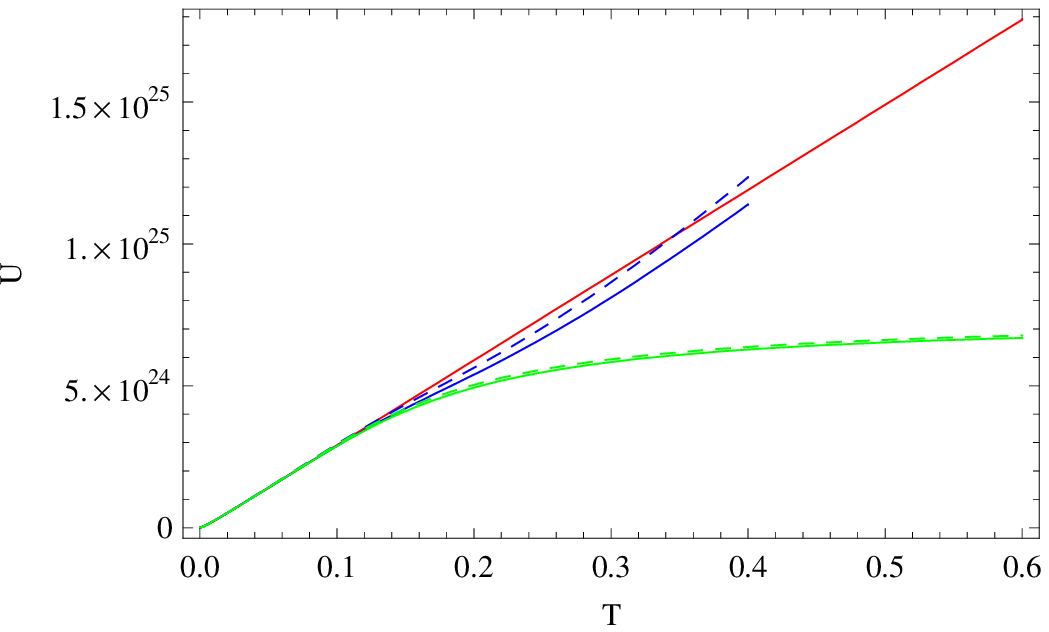}}\\
  \scalebox{0.75}{\includegraphics{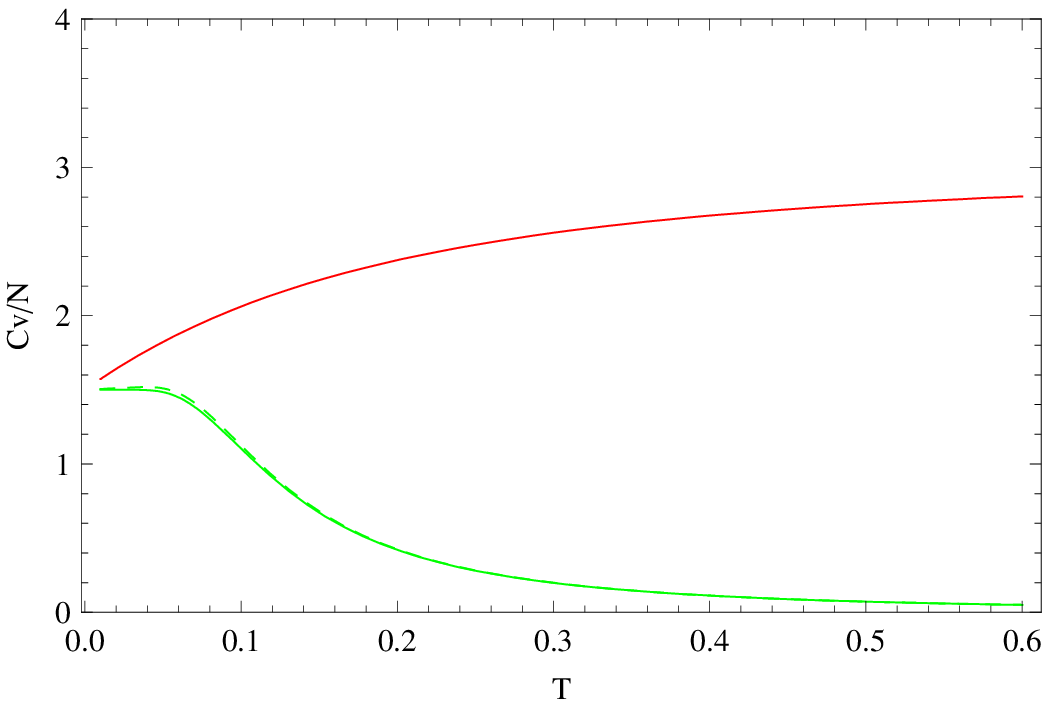}}
  \scalebox{0.75}{\includegraphics{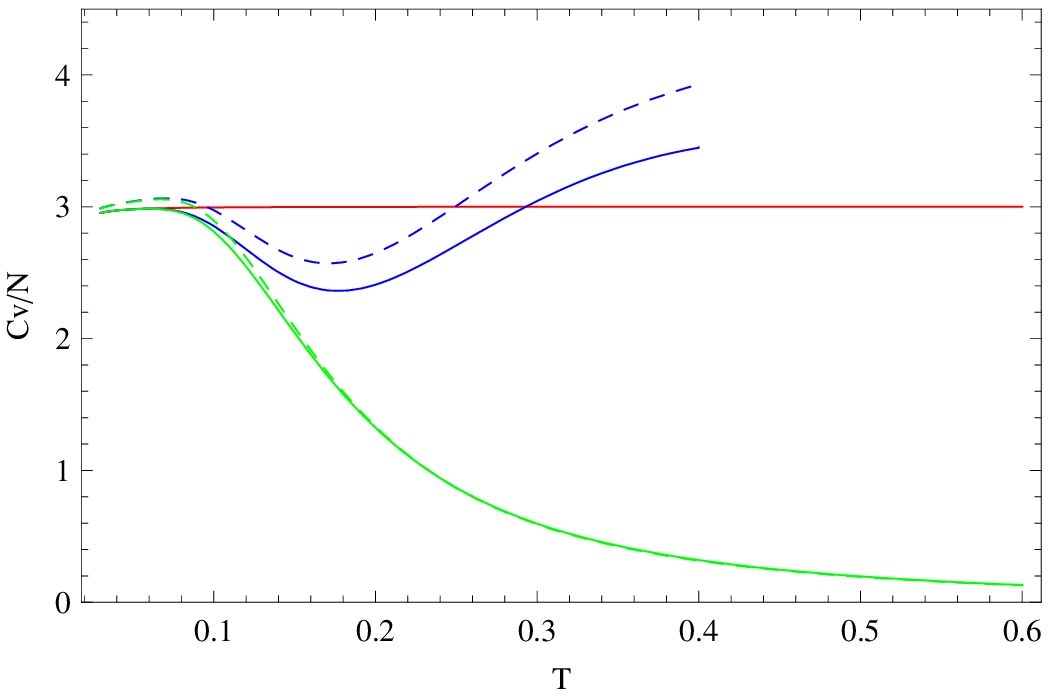}}\\
  \scalebox{0.75}{\includegraphics{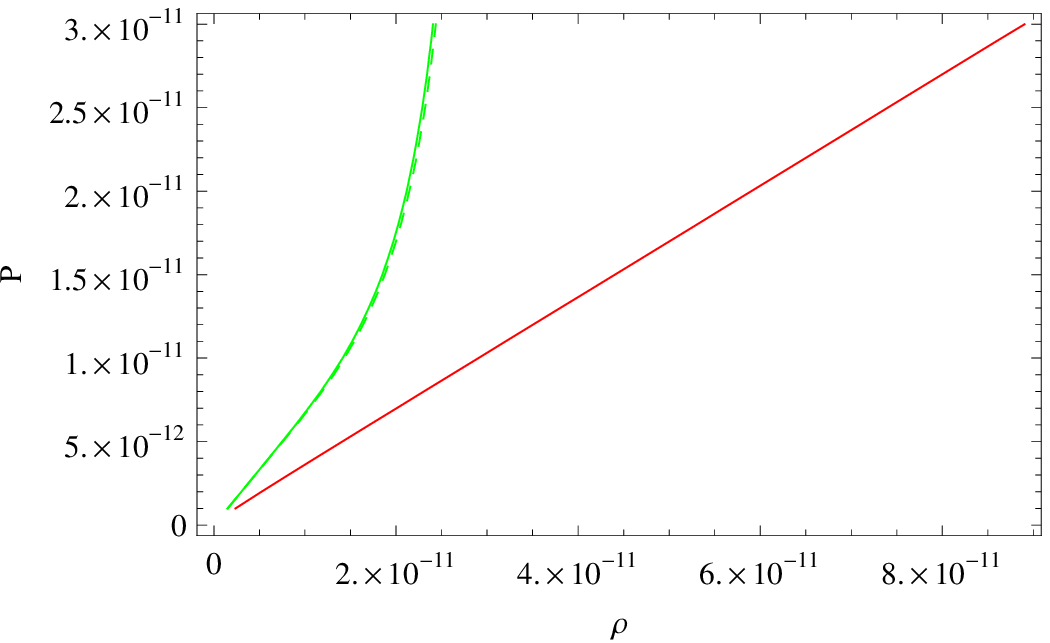}}
  \scalebox{0.75}{\includegraphics{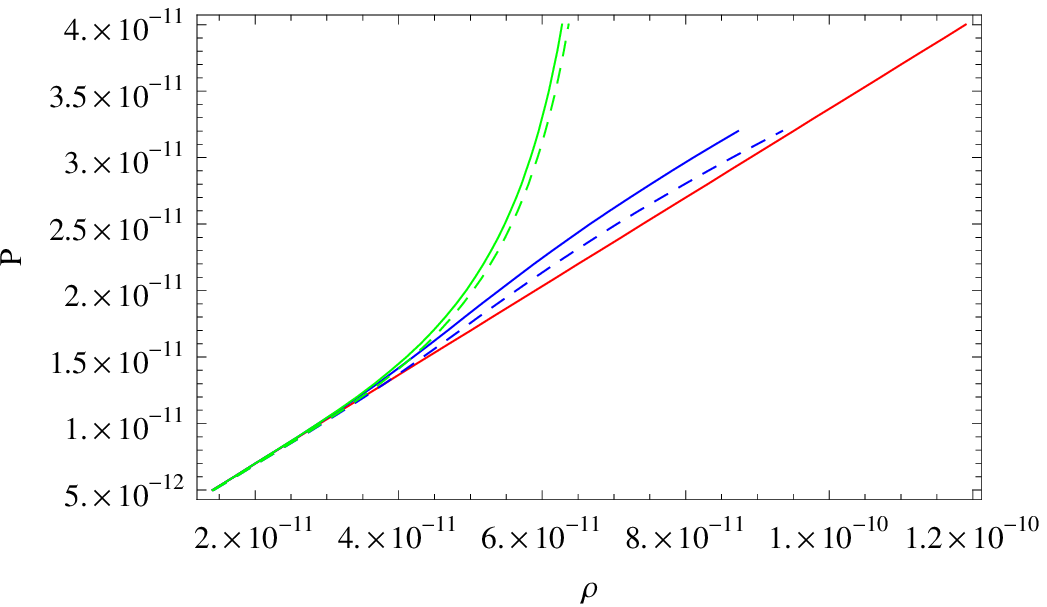}}\\
\end{center}
\caption{The plots for the unmodified measure are in solid lines while those for modified measure are in dashed lines.
The SR plots are shown in red. The figures in the left column are for Case I while those in the right column are for Case III. The DSR plots for Case I are plotted in green. For Case III, the leading order behaviours are plotted in blue. The numerical plots for Case III are also shown in green for comparison. Different values for the parameters in natural units are as follows: $V=10^{35}, N=10^{25},\kappa=1, m_{0} = 1 \, {\rm (Case \,\, I)}, 0.01 \, {\rm (Case\,\, III)}, a_{0}=1, a_{1}=0.2, a_{2}=a_{3}=...=0$.
}
\label{fg.TQ}
\end{figure}
%
%
The above quantities for Case I can be found by using (\ref{Zcase2}) and they have been plotted in Fig \ref{fg.TQ}. Now we shall obtain the leading order thermodynamics for Case III.
If we denote  the free energy, pressure, entropy, internal energy, internal energy density and heat capacity obtained in the SR or massless cases by $F_0, P_0, S_0, U_0$, $\rho_0$ and $C_{V0}$ respectively, and write $Z_1=Z_{10}+Z_{1corr}$, where $\l Z_{10},Z_{1corr}\r=\l Z_{1SR},Z_{1SRcorr}\r$ or $\l Z_{1ml},Z_{1mlcorr}\r$, we have
\begin{eqnarray}
F&=&F_{0} - {N\over \beta}ln \left(1+{Z_{1corr}\over Z_{10}}\right)=F_{0}-{N\over \beta} {Z_{1corr}\over Z_{10}} + \mathcal{O}\l\left({Z_{1corr}\over Z_{10}}\right)^2\r, \\
S&=&N\left[\ln\left({Z_1\over N}\right)+1\right]-{\beta N\over Z_1}{\partial Z_1\over \partial \beta}\nonumber\\
&=&S_{0}+{NZ_{1corr}\over Z_{10}}-{\beta N\over Z_{10}}
\left({\partial Z_{1corr}\over \partial \beta}-{Z_{1corr}\over Z_{10}}{\partial Z_{10}\over \partial \beta}\right)+ \mathcal{O}\l\left({Z_{1corr}\over Z_{10}}\right)^2\r, \\
U&=&U_{0}\l1-{Z_{1corr}\over Z_{10}}\r-{N\over Z_{10}}{\partial Z_{1corr}\over \partial \beta}+ \mathcal{O}\l\left({Z_{1corr}\over Z_{10}}\right)^2\r, \\
C_V&=&C_{V0}\l1-\frac{Z_{1corr}}{Z_{10}}\r-\beta^2\left[-
{U_{0}\over Z_{10}}{\partial Z_{1corr}\over \partial \beta}+{Z_{1corr}\over Z_{10}^2}U_{0}{\partial Z_{10}\over \partial \beta}+
{N\over Z_{10}^2}{\partial Z_{10}\over \partial \beta}{\partial Z_{1corr}\over \partial \beta}-{N\over Z_{10}}{\partial^2 Z_{1corr}\over \partial^2 \beta}\right]\nonumber\\
&&+ \mathcal{O}\l\left({Z_{1corr}\over Z_{10}}\right)^2\r, \\
\rho&=&\rho_{0}\left(1-{Z_{1corr}\over Z_{10}}\right)-{n\over Z_{10}}{\partial Z_{1corr}\over \partial \beta}+ \mathcal{O}\l\left({Z_{1corr}\over Z_{10}}\right)^2\r,
\end{eqnarray}
where $n = \frac{N}{V}$ is the number density.
The correction to $F$ depends on the ratio of $Z_{1corr}$ and $Z_{10}$ and is independent of the volume $V$ of the system. Note that this is true to all orders.
Hence, the pressure $P$ of the system which is defined as $ P=-\left({\partial F\over \partial V}\right)_{N,T}$ gets no correction:
\begin{equation}
  P=P_0.
\end{equation}
The equation of state in SR is (see $(8.128)$ and $(8.134)$ of \cite{greinerbook}) 
\beq
P_{SR}={\rho_{SR}\over3-\beta m+\frac{K_1(\beta m)}{K_2(\beta m)}\beta m},
\eeq 
which gives the following DSR equation of state
\begin{eqnarray}
 P&=&{\left(\rho+{Z_{1SRcorr}\over Z_{1SR}}\rho+{n\over Z_{1SR}}{\partial Z_{1SRcorr}\over \partial \beta}\right)\over3-\beta m+\frac{K_1(\beta m)}{K_2(\beta m)}\beta m}+ \mathcal{O}\l\left({Z_{1SRcorr}\over Z_{1SR}}\right)^2\r.
\end{eqnarray}
In Fig \ref{fg.TQ}, we have plotted the various thermodynamic quantities for Cases I and III as a function of $T$ and compared them with the SR case. The $P$ vs $\rho$ plots have been obtained by varying $T$ keeping all other parameters fixed. The qualitative natures of the plots for different cases are same. In case of $S$ there are two competing effects: while the  cutoff tries to reduce $S$ by limiting the number of accessible states, the modified dispersion tries to increase $S$ by enhancing the Boltzmann weight $\exp(-\e/T)$ (note that $\e_{DSR}(p)< \e_{SR}(p)$ for a given momentum state $p$, the change being more for greater value of the parameter $m$). At low temperatures, the latter is dominant and $S_{DSR}> S_{SR}$. For our choice of parameters this is clearly visible for the plot of $S$ in Case I. In the high $T$ regime, the cutoff effect comes into play and  $S_{DSR}< S_{SR}$. The cutoff also saturates $U$ as $T$ increases, and $C_V \rightarrow 0$, resulting in a steeper equation of state. Here we make an interesting observation. There have been attempts to define velocity in DSR \cite{Kosinski:2002gu}. If we adopt the usual definition for the speed of sound $c_s = \sqrt{\frac{\partial P}{\partial \rho}}$, then we observe that $c_s$ grows without any bound. Possibility of such scenarios has been discussed in \cite{Kim:2004uj}.

For given choice of parameters in case of quantities like $S$ and $U$, 
the leading order behaviours for Case III match with the numerical plots up to $T \sim 0.15$, 
while in case of $C_V$ which contains second order derivatives of the partition function with respect to $T$, 
the leading order behaviours match with the numerical plots up to $T \sim 0.09$. 
Note that the leading order behaviours have been obtained assuming $\frac{1}{\beta m_0}$ to be finite and $\frac{m_0}{\kappa}\rightarrow 0$ 
which in turn implies $\frac{1}{\beta\kappa}=\frac{T}{\kappa}\rightarrow 0$. Hence as $T$ increases, the leading order plots 
depart from their numerical counterparts. 

We conclude with a summary of our investigation of the ideal gas thermodynamics in DSR framework.
In this paper we have used the dispersion relation (\ref{MS}) and have considered the modified phase space measure (the modification being isotropic and expandable in Taylor series).
We consider three cases separately ($m=\kappa, m>\kappa, m<\kappa$). 
The single particle partition function has been shown to be smooth in $m_0 \in (0,\kappa)$ (see Appendix \ref{app1}).
For the case $m=\kappa$, a simple analytical form for the partition function is obtained (see (\ref{Zcase2})) while a series solution for the partition function has been obtained for $m<\kappa$ (see (\ref{Z}) and (\ref{kstar_mr})).
In doing so, new type of special functions (Incomplete Modified Bessel functions) emerge.
We observe that DSR thermodynamics is non-perturbative in the SR and massless limits.
Using the leading order solutions, we derive thermodynamic quantities like the 
free energy, pressure, entropy, internal energy and heat capacity (see Fig.\ref{fg.TQ}).
A stiffer equation of state is found.

\section{Acknowledgement}
We would like to thank Amit Samanta for valuable discussions, in particular for his inputs in Appendix \ref{app1}.
We would also like to thank Apoorva D Patel and Diptiman Sen for their useful comments.
We further thank the PRD referee for his suggestions and comments.

\section*{Appendices}
\appendix
\section{Convergence of $K^*\l x,y\r$}\label{kstar_convergence}
$K^*\l x,y\r$ given by (\ref{kstar_mr},\ref{m0},\ref{m1},\ref{mr_series}) is convergent if the following two series are convergent:
\begin{equation}
 S=\sum_{r=2}^{\infty}t_r\frac{(-x)^{2r-3}}{(2r-3)!}
\end{equation}
and
\begin{equation}
 S^\prime=\sum_{r=2}^{\infty}t_r\sum_{k=1}^{2r-3}\frac{(-x)^{k-1}}{(2r-3)(2r-4)...(2r-2-k)}\left(\frac{1}{y}\right)^{2r-2-k}.
\end{equation}
$S$ can be easily proved to be absolutely convergent using Ratio test.
For $S^\prime$ first consider the following double series:
\begin{equation}
 S^{\prime\prime}=\sum_{r=2}^{\infty}t_r\sum_{k=1}^{\infty}\frac{(-x)^{k-1}}{(2r-3)(2r-4)...(2r-2-k)}\left(\frac{1}{y}\right)^{2r-2-k}
=\sum_{r=2}^{\infty}\sum_{k=1}^{\infty}a_{r,k}.
\end{equation}
Let us first test the convergence of $S^{\prime\prime}$ (See theorem (2.7) of \cite{balmohan}).
The row series $S_r$ (for a fixed $r$) and the column series $S_k$ (for a fixed $k$) are defined as
\begin{equation}
 S_{r}=\sum_{k=1}^{\infty}a_{r,k},
\end{equation}
\begin{equation}
 S_{k}=\sum_{r=2}^{\infty}a_{r,k}.
\end{equation}
The ratio tests for $S_r$ and $S_k$ show that they are absolutely convergent (for $y>1$).
Also $\displaystyle{\lim_{r,k\rightarrow \infty}}\left|\frac{a_{r,k+1}}{a_{r,k}}\right|=0<1$. Hence, $S^{\prime\prime}$ is absolutely convergent.
Now
\begin{equation}
 |S^{\prime\prime}|=\sum_{r=2}^{\infty}\sum_{k=1}^{\infty}|a_{r,k}|=\sum_{r=2}^{\infty}\sum_{k=1}^{2r-3}|a_{r,k}|+L \quad \quad \quad ;L\geq 0.
\end{equation}
As $|S^{\prime\prime}|$ is convergent (or in other words $S^{\prime \prime}$ is absolutely convergent) 
we must have $\left|S^{\prime}\right|=\displaystyle{\sum_{r=2}^{\infty}\sum_{k=1}^{2r-3}}|a_{r,k}|$ to be convergent (or in other words $S^\prime$ to be absolutely convergent).
Thus the series expansion of $K^*\l x,y\r$ is absolutely convergent.
\section{Continuity and differentiability of the partition function in $m_0$}
\label{app1}
We shall show that $Z_1^0$ is continuous in $m_0$ for $m_0\in [0,\kappa]$.
After integrating over the angular coordinates (\ref{z10}) gives
\begin{equation}
Z_1^0\l m_0 \r = \frac{2V}{\l2\pi\r^2}\int_{0}^{\kappa}dp \,\,\, p^2 e^{-\beta[\e(p,m_0)-m_0]} 
=\frac{2V}{\l2\pi\r^2}\int_{0}^{\kappa}dp \,\,\, f(p,m_0).
\label{Z10_m0}
\end{equation}
The integrand $f(p,m_0) = p^2 e^{-\beta[\e(p,m_0)-m_0]}$ is a continuous bounded function of 
$p$ and $m_0$ in the range $m_0\in [0,\kappa], p\in [0,\kappa]$.
Thus $Z_1^0\l m_0 \r$ is a continuous function of $m_0$ as the function $g(p)=C$,
where $C$ is the upper bound of $|f(p,m_0)|$ in the range $m_0\in [0,\kappa], p\in [0,\kappa]$, satisfies
$g(p)\geq |f(p,m_0)|$ for all $m_0\in [0,\kappa], p\in [0,\kappa]$ and is integrable as
 $\displaystyle{\int_{0}^{\kappa}}dp\,\,\, g(p) = C\kappa<\infty$ (see Lemma 1 in $\S$ 5.12 of \cite{fleming}).\\
\indent The derivative of the integrand with respect to $m_0$ is given by
\begin{equation}
 \frac{\partial f(p,m_0)}{\partial m_0} = \beta f(p,m_0) 
\l1-\frac{\e^2(p,m_0)-p^2}{m_0 \l1-\frac{2m_0}{\kappa}\r^{1/2}\l p^2-\frac{m_0^2}{\frac{2m_0}{\kappa}-1}\r^{1/2}}\r.
\end{equation}
It has 2 poles (and also branch points) in the complex $p$-plane at $p=\pm \frac{m_0}{\l\frac{2m_0}{\kappa}-1\r^{1/2}}$.
We note that the poles and the branch points remain to be at the same positions for all higher order derivatives of $f(p,m_0)$
with respect to $m_0$. 
For $m_0=0$ both the poles are at $p=0$ and as $m_0$ increases the poles separate towards the imaginary axis.
They keep on moving on the imaginary axis till they reach $\pm i \infty$ at $m_0=\frac{\kappa}{2}$.
After that they start to come closer to each other on the real line and finally at $m_0=\kappa$ they stop at $p=\pm \kappa$.
Note that for all $m_0\in(0,\kappa)$ the poles are never on the contour of integration (the real line from $p=0$ to $p=\kappa$)
and the functions $\l\frac{\partial f(p,m_0)}{\partial m_0},\frac{\partial^2 f(p,m_0)}{\partial m_0^2}, etc.\r$ 
remain to be bounded.
This (by the same argument as given in the case of the continuity of $Z_1^0\l m_0 \r$) ensures the infinite-order differentiability
of $Z_1^0\l m_0 \r$ in $m_0\in(0,\kappa)$ and the derivatives can be found by using the Leibniz rule 
(see Lemma 2 in $\S$ 5.12 of \cite{fleming}). Note that the fact that we are not being able to say about the differentiability 
of $Z_1^0$ at $m_0=0$ could be a relic of the non-analytic part in (\ref{zm0}).

\end{document}